\documentclass[a4paper,onecolumn,11pt,unpublished]{quantumarticle}
\pdfoutput=1
\usepackage[utf8]{inputenc}
\usepackage[english]{babel}
\usepackage[T1]{fontenc}
\usepackage{amsmath}
\usepackage{hyperref}
\usepackage[numbers]{natbib}
\usepackage{tikz}
\usepackage{lipsum}
\usepackage{subfigure}
\usepackage{bm}
\usepackage{amssymb}
\usepackage{amsthm}
\usepackage{multirow}
\usepackage{multicol}
\usepackage[ruled]{algorithm2e}

\newcommand{\respred}[1]{#1}

\newtheorem{example}{Example}

\begin{document}

\title{Non-Markovian Quantum Gate Set Tomography}

\author[1,3,4]{Ze-Tong Li}
\author[1,3,4]{Cong-Cong Zheng}
\author[5]{Fan-Xu Meng}
\author[2,3,4]{Han Zeng}
\author[6]{Tian Luan}
\author[2,3,4,7]{Zai-Chen Zhang}
\author[1,3,4,7]{Xu-Tao Yu}
\email{yuxutao@seu.edu.cn}

\affil[1]{State Key Laboratory of Millimeter Waves, Southeast University, Nanjing 210096, China}
\affil[2]{National Mobile Communications Research Laboratory, Southeast University, Nanjing 210096, China.}
\affil[3]{Frontiers Science Center for Mobile Information Communication and Security, Southeast University, Nanjing 210096, China.}
\affil[4]{Quantum Information Center, Southeast University, Nanjing 210096, China.}
\affil[5]{College of Artificial Intelligence, Nanjing Tech University, Nanjing 211800, China}
\affil[6]{Yangtze Delta Region Industrial Innovation Center of Quantum and Information Technology, Suzhou 215100, China.}
\affil[7]{Purple Mountain Lab, Nanjing 211111, China.}

\maketitle

\begin{abstract}
  Engineering quantum devices requires reliable characterization of the quantum system, including qubits, quantum operations (also known as instruments) and the quantum noise. Recently, quantum gate set tomography (GST) has emerged as a powerful technique for self-consistently describing quantum states, gates, and measurements. However, non-Markovian correlations between the quantum system and environment impact the reliability of GST. To address this, we propose a self-consistent operational framework called instrument set tomography (IST) for non-Markovian GST. Based on the stochastic quantum process, the instrument set describes instruments and system-environment (SE) correlations. We introduce a linear inversion IST (LIST) to describe instruments and SE correlations without physical constraints. The disharmony of linear relationships between instruments is detected. Furthermore, we propose a physically constrained statistical method based on the maximum likelihood estimation for IST (MLE-IST) {with adjustable dimensions}. MLE-IST shows significant flexibility in adapting to different types of devices, such as noisy intermediate-scale quantum (NISQ) devices, by adjusting the model and constraints. Experimental results demonstrate the effectiveness and necessity of simultaneously describing instruments and SE correlations. Specifically, the LIST and MLE-IST obtains significant improvement on average square error reduction in the imperfect implemented simulations by orders of -23.77 and -6.21, respectively, compared to their comparative methods. {Remarkably, real-chip experiments indicate that a polynomial number of parameters with respect to the Markovian order are sufficient to characterize non-Markovian quantum noise in current NISQ devices.} Consequently, IST provides an essential and self-consistent framework for characterizing, benchmarking, and developing quantum devices in terms of the instrument set.
\end{abstract}

\section{Introduction}
Quantum computing necessitates engineering reliable and controllable quantum devices that can accurately manipulate quantum states. Unfortunately, the current generation of quantum devices, also known as noisy intermediate-scale quantum (NISQ), is susceptible to non-ignorable quantum noise resulting from imperfect implementations of quantum gates and system-environment (SE) correlations \cite{papic2023Error}. The characterization of qubits, operations, and entire processors becomes crucial in analyzing the impact of quantum noise. This process, known as a key aspect of quantum characterization, verification, and validation (QCVV), provides essential insights for device manufacturing and calibration.

Based on different assumptions, numerous protocols have been proposed for the task under common skeleton of quantum tomography \cite{banaszek2013Focus,smolin2012Efficient,blume-kohout2010Optimal,koutny2022Neuralnetwork,riebe2006Process,mohseni2008Quantumprocess,surawy-stepney2022Projected,greenbaum2015Introduction,nielsen2021Gate}: (1) preparing a set of experiments consisting of quantum states, circuits and measurements; (2) collecting data by executing the prepared experiments; (3) performing estimation algorithms to derive the desired results of quantum states, processes, and/or measurements. Gate set tomography (GST) is the most comprehensive method among these tomographic approaches, enabling the operational and self-consistent characterization of quantum gates, state preparations, and measurements (SPAM) without any prior knowledge requirements about the experiment components \cite{greenbaum2015Introduction,nielsen2021Gate}. Conversely, quantum state tomography (QST) \cite{banaszek2013Focus,smolin2012Efficient,blume-kohout2010Optimal,koutny2022Neuralnetwork} and quantum process tomography (QPT) \cite{riebe2006Process,mohseni2008Quantumprocess,surawy-stepney2022Projected} generally require the complete knowledge of non-target parts in the experiments. The GST successfully describes two-time noisy quantum gates by completely positive trace-preserving (CPTP) maps under the Markovian assumption. 

However, no system is isolated \cite{pollock2018NonMarkovian}. Sufficient evidence suggests that non-Markovian multiple time correlations significantly impact current generation of quantum devices \cite{blume-kohout2017Demonstration,proctor2022Measuring,white2020Demonstration,sarovar2020Detecting}. Non-Markovian correlations, where past operations influence the behavior of current operations and can result in the theoretical violation of completely positive trace-preserving (CPTP) constraints \cite{proctor2022Measuring,milz2021Quantum}, not only disrupt tomography under the Markovian model, but also the degradation or vanishment of the effectiveness of quantum error-correcting codes \cite{nickerson2019Analysing,clader2021Impact}. Therefore, Markovian two-time CPTP maps are not sufficient to describe entire dynamics of the quantum device. Correlations across multiple time scales should be considered in the device characterization. 

Based on the quantum stochastic process that represents multiple time correlations \cite{milz2021Quantum}, non-Markovian open quantum dynamics can be modeled by the system, environment, instruments act on the system, and unitaries act on the system and environment simultaneously. From the perspective of an experimenter, only the instruments representing interventions on the system, such as quantum gates and measurements, are accessible. These instruments can be represented by completely positive (CP) and trace-non-increasing (TNI) maps. Aiming at operationally describing the time-dependent SE correlations, process tensor tomography (PTT) \cite{pollock2018NonMarkovian,guo2022Reconstructing,milz2018Reconstructing,white2022NonMarkovian} relaxes the Markovian constraints to perform the non-Markovian quantum process tomography. It constructs a well-defined CP process tensor with unit trace by interventions of a set of known instruments that can span the space of bounded linear operators on system quantum state (also known as informationally complete). However, differences between the theoretical knowledge and the practical performance of these instruments may disturb the reconstruction of the process tensor \cite{white2022NonMarkovian}. A simple example is that PTT may generate inconsistent two process tensors using two set of imperfect informationally complete instruments. Consequently, the characterization of real quantum devices requires a self-consistent method that can tomographically describe both non-Markovian SE correlations and faulty instruments, which is the direct motivation for this work.

To address these issues, we propose a self-consistent framework called the instrument set tomography (IST) for performing GST on non-Markovian open quantum systems. We define full and reduced instrument sets based on the quantum stochastic process. Specifically, the full instrument set explicitly consists of instruments, SE unitaries, and the initial SE state. The reduced instrument set utilizes the process tensor to represent inaccessible SE correlations which consists of SE unitaries and the initial SE state. We first propose the linear inversion IST (LIST), which is a simple, closed-form method based on the reduced instruments set. LIST allows us to simultaneously estimate both the instruments and the process tensor. The disharmony of the linear relationship of available instruments can be detected, when the instruments at a time step are not linear independent. 
Moreover, the gauge optimization is required at the end of LIST because of the gauge freedom similar to that in the GST. Although the estimated results may not satisfy the physical constraints due to the lack of constraints in the gauge optimization, they are consistent with the probability measurement data.

Then, we propose a statistical IST method based on the maximum likelihood estimation (MLE) to extract additional information from overcomplete measurement data. The MLE-IST models the instruments and SE correlations in a flexible way accommodating various assumptions. By introducing constraints, the method ensures results to be physical. Furthermore, it enables the explicit and feasible estimation of SE unitaries based on the full instrument set instead of the process tensor.

Particularly, we also demonstrate how to implement IST on NISQ devices. The effectiveness of methods is measured by the square error of probabilities (SEP). Comparing with corresponding Markovian tomographic schemes, the LIST and MLE-IST obtains significant average SEP reduction by order of -23.40 and -6.00 in simulations, respectively. We also implement the IST on real quantum devices, where the LIST achieves SEP reduction by order of -12.75, and the MLE-IST gains the average improvement on SEP of 57.2\%. Consequently, the experimental results indicate the effectiveness and indispensability of simultaneously characterizing instruments and non-Markovian correlations. The IST provides an essential, self-consistent, and reliable method for benchmarking and developing quantum devices under non-Markovian situations in the aspect of instrument set. 

This paper is organized as follows. We briefly review the quantum stochastic process and define the instrument set in Section~\ref{sec:instrument_set}. Then, the framework of IST including LIST and MLE-IST is proposed in Section~\ref{sec:IST}. Based on the characteristics of current quantum devices, we introduce how to implement IST on NISQ devices in Section~\ref{sec:nisq_ist}. Results of simulations and real-chip experiments are demonstrated in Section~\ref{sec:result}. Finally, we conclude this work in Section~\ref{sec:discussion}.

\section{Quantum Stochastic Process and Instrument Set}\label{sec:instrument_set}

Before presenting the IST, we recall the quantum stochastic process \cite{milz2021Quantum} describing open quantum dynamics. Then, full and reduced definitions of instrument set are given. 

Consider a $d$-dimensional quantum system $\rho_S\in\mathcal{H}_S$ that is subject to $k$-time-step interventions suffering non-Markovian correlations with the environment. The experimenter intervenes the quantum system at time step $t$ by a CPTNI instrument from $t$-available set
\begin{align}
  \mathcal{J}^{(t)} := \left\{\mathcal{A}^{(t)}_{0}, \mathcal{A}^{(t)}_{1},\dots, \mathcal{A}^{(t)}_{m_t-1}\right\},
\end{align}
where $\mathcal{A}^{(t)}_{x_t}$ are bounded operators on the $\mathcal{H}_S$, $t=1,\dots,k$, and $m_t$ is the number of available instruments at time step $t$. Note that available sets may vary over time steps. \respred{Each intervention of an instrument transforms the input quantum state into the output quantum state and/or outputs a value. For example, an ideal quantum gate is an instrument that transforms the input quantum state as a unitary operator with nothing output. Furthermore, a positive operator-valued measurement (POVM) is an instrument that projects the quantum state into an eigenstate of the POVM operator and outputs the eigenvalue. This concept of instrument fits both superconductive and photonic quantum computers.}

\begin{figure*}[t]
  \centering
	\includegraphics[width=0.8\textwidth]{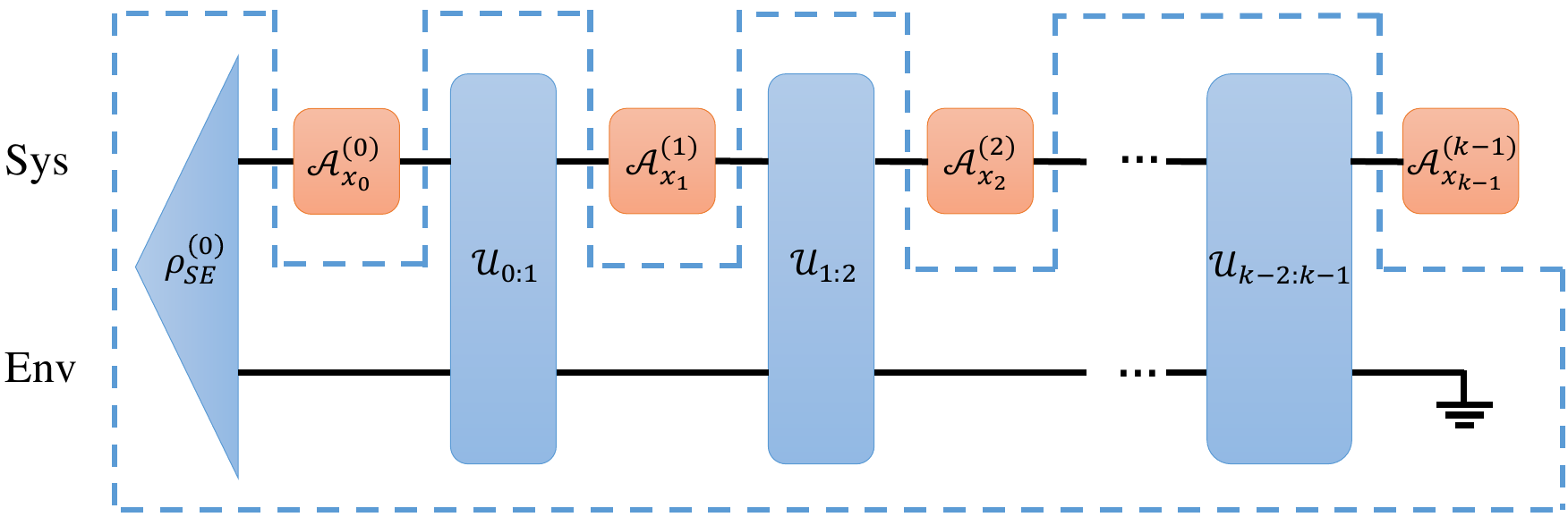}
  \caption{\label{fig:qsp} Model of operational open quantum process. $\rho_{SE}^{(0)}$ represents the initial state, where the system (Sys) and environment (Env) can be entangled. Instruments are represented by red blocks, which are the only accessible part to an experimenter. There is a unitary representing the SE unitary dynamics acting on both the system and environment dimensions between two adjacent time steps. The blue dashed line separates the accessible and inaccessible parts for an experimenter.
  }
\end{figure*}

The non-Markovianity of a quantum system implies the existence of correlations spanning multiple time steps between the system and environment. These correlations enable the state of the environment to be alternated by instrument interventions on the system and information to be transferred by the environment. The quantum stochastic process theorem \cite{milz2021Quantum} reveals that the non-Markovian open quantum system, as depicted in Figure~\ref{fig:qsp} \cite{pollock2018NonMarkovian}, can be operationally described by a quantum state $\rho_{SE}\in \mathcal{H}_{SE}:=\mathcal{H}_{S}\otimes \mathcal{H}_{E}$ consisting of the system and a $d^\prime$-dimensional environment $\rho_E\in \mathcal{H}_{E}$ with instrument interventions on $\mathcal{H}_{S}$ and SE unitary evolutions on $\mathcal{H}_{SE}$, {where $d^\prime$ is exponentially large to $k$ to represent arbitrary open quantum dynamics}. Specifically, the operational non-Markovian open quantum system evolves starting with a pure SE state $\rho_{SE}^{(0)}\in \mathcal{H}_{SE}$ before time step $0$. At time step $t$, the experimenter chooses an instrument $\mathcal{A}^{(t)}_{x_t}$ from $t$-available set to intervene the system and get the output of $\mathcal{A}^{(t)}_{x_t}$. Subsequently, the SE state is evolved by the SE unitary evolution described by $\mathcal{U}_{t:t+1}$ to be $\rho_{SE}^{(t+1)} = \mathcal{U}_{t:t+1}\circ\left(\mathcal{A}^{(t)}_{x_t}\otimes \mathcal{I}\right)\!\left(\rho_{SE}^{(t)}\right)$.
Hence, the probability of performing instruments labeled by $\bm{x}$ is
\begin{equation}\label{eq:se_evo_prob_tr}
  p_{\bm{x}}=\mathrm{Tr}\left[\bar{\mathcal{A}}^{(k-1)}_{x_{k-1}}\bigcirc_{t=0}^{k-2}\left(\mathcal{U}_{t:t+1}\circ\bar{\mathcal{A}}^{(t)}_{x_t}\right)\left(\rho_{SE}^{(0)}\right)\right],
\end{equation}
where $\bar{\bullet}:= \bullet \otimes \mathcal{I}$. 

Note that in a non-Markovian open quantum system, there is a boundary between parts accessible and inaccessible to an experimenter. Specifically, an experimenter cannot access the quantum state directly. All information of the quantum state the experimenter obtained should be with the help of instrument interventions. Generally, experimenters are interested in the probability of $\bm{x}$, which leads to the requirement of instrument set that is sufficient to describe the probability when an experimenter chooses an arbitrary sequence of instruments from available sets.

Equation \eqref{eq:se_evo_prob_tr} shows that the probability can be determined when the instruments, the SE unitaries, and the initial state are given. Therefore, the instrument set describing the operational open quantum dynamics of the device can be defined as
\begin{align}\label{eq:instrument_set_full}
  \mathfrak{I}_{\mathrm{full}} := \left\{\mathcal{J}, \mathcal{U},\rho^{(0)}_{SE}\right\},
\end{align}
where $ \mathcal{J}:=\left\{\mathcal{J}^{(t)}\right\}_{t=0}^{k-1}$ and $\mathcal{U}:=\left\{\mathcal{U}_{t:t+1}\right\}_{t=0}^{k-2}$.
This full definition explicitly depends on the inaccessible initial state and the SE unitaries between time steps in which the experimenter may be interested. However, explicit characterization of the initial state and SE unitaries is difficult. Due to the freedom of environment, the initial state and SE unitaries are non-unique without violation of a complete set of probabilities even if instruments are perfectly known. This can be easily verified by inserting a unitary and its inverse on environment only at a time step, cf. Example~\ref{eg:nunique}. Hence, the tomography of the full instrument set always determines a possible result from a cluster of feasible results.

By utilizing the process tensor $\mathcal{T}$ representing the inaccessible parts \cite{pollock2018NonMarkovian,milz2021Quantum,white2022NonMarkovian}, the probability to performing instruments labeled by $\bm{x}$ can be described as 
\begin{align}\label{eq:pt_def}
  p_{\bm{x}} = \mathcal{T}\left(\mathcal{A}^{(0)}_{x_0},\dots,\mathcal{A}^{(k-1)}_{x_{k-1}}\right),
\end{align}
which implies the sufficiency to determine the measurement probability by given $\mathcal{T}$ and $\mathcal{J}$. Therefore, the reduced instrument set can be defined as 
\begin{align}\label{eq:instrument_set_reduced}
  \mathfrak{I}_{\mathrm{reduced}}:=\left\{\mathcal{J}, \mathcal{T}\right\}.
\end{align}

In the following, we consistently utilize the Pauli transfer matrix (PTM) representation to describe instruments, quantum states, and process tensors. Specifically, we denote the PTM of instrument $\mathcal{A}$ as $A$ and the PTM of quantum state $\rho$ as $\vert \rho\rangle\!\rangle$. Furthermore, the PTM representation of the process tensor $\Upsilon_{\mathcal{T}}$ is defined as
\begin{align}\label{eq:se_evo_prob_pt}
  p_{\bm{x}} = \mathrm{Tr}\left[\Upsilon_{\mathcal{T}}^\dagger\left(A^{(0)}_{x_0}\otimes\dots\otimes A^{(k-1)}_{x_{k-1}}\right)\right]
\end{align}
instead of directly applying the Choi-Jamiołkowski isomorphism (CJI) representation for the sake of notation consistency. 
Note that the effects of imperfect instrument implementation can be integrated into the representation of both $A^{(t)}_{x_t}$ and $\Upsilon_{\mathcal{T}}$, as shown in Example~\ref{eg:nunique_reduced}, which also leads to the non-uniqueness of tomographic result.

These two definitions of instrument sets will be exploited to propose the IST with clear declaration. The non-uniqueness of both full and reduced instrument sets in terms of probabilities bears a resemblance to the gauge freedom observed in conventional GST, but their intrinsic sources differ and will be elucidated in Section~\ref{sec:IST}. Nonetheless, we still refer to the non-uniqueness in the IST as gauge freedom for conceptual coherence.


\section{Instrument Set Tomography} \label{sec:IST}

In this section, we propose the IST for non-Markovian GST. First, we introduce a linear inversion IST (LIST) based on the reduced instrument set. Then, we propose a statistical IST method based on maximum likelihood estimation, using the full instrument set. The requirement of informationally completeness of available set at each time step is relaxed. This is because the IST takes the quantum device as input, and it may not equip an informationally complete available set at a specific time step. For example, current superconductive NISQ devices can only perform limited types of quantum gates subject to noisy unitaries evolutions. The informationally completeness cannot be guaranteed due to the uncontrollable noise. Moreover, there are NISQ devices whose available sets cannot span $SU(d)$ when a time step is specified as a time slot for performing a single quantum gate \cite{ibmquantum2021Ibm_perth}. In such cases, the IST should still provide meaningful results. Consequently, the IST always determines a result that is feasible within the space spanned by the quantum device.

\subsection{Linear Inversion IST}\label{subsec:LIST}

Using the Pauli transfer matrix (PTM) representation defined in \cite{greenbaum2015Introduction}, the probability of performing instruments labeled by $\bm{x}$ as described in Eq.\eqref{eq:se_evo_prob_tr} can be reformulated as
\begin{align}
  p_{\bm{x}} =& \langle\!\langle0_{SE}\vert \bar{A}^{(k-1)}_{x_{k-1}}\prod_{t=0}^{k-2}U_{t:t+1}\bar{A}^{(t)}_{x_t} \vert \rho^{(0)}_{SE}\rangle\!\rangle,\label{eq:nm_exp_ptm}
\end{align}
where $\vert \bullet \rangle\!\rangle$ and $\langle\!\langle \bullet\vert$ represent superoperators of a quantum state and a positive operator-valued measurement (POVM) operator, respectively. Additionally, ${A}^{(t)}_{x_t}$ and $U_{t:t+1}$ are the PTM representations of ${\mathcal{A}}^{(t)}_{x_t}$ and $\mathcal{U}_{t:t+1}$, respectively.

Focusing on the time step $t$, let $\bm{x}^+$ and $\bm{x}^-$ represent the indexes before and after time step $t$ in a $k$-time-step experiment, respectively. Reforming the probability as 
\begin{align}
  p_{\bm{x^+}\bm{x^-}}(x_t)
  =\mathrm{Tr}\left[\!\sum_{ij}\langle\!\langle F^E_{\bm{x}^+\!,i}\vert F^E_{\bm{x}^-\!,j}\rangle\!\rangle \vert F^S_{\bm{x}^-\!,j}\rangle\!\rangle\langle\!\langle F^S_{\bm{x}^+\!,i}\vert A^{(t)}_{x_t}\!\right],
\end{align}
where 
\begin{align}
  &\langle\!\langle F^{SE}_{\bm{x}^+}\vert = \sum_{i} \langle\!\langle F^S_{\bm{x}^+\!,i}\vert \otimes \langle\!\langle F^E_{\bm{x}^+\!,i}\vert =\langle\!\langle 0_{SE}\vert \left[\prod_{r=1}^{k-t-2}\bar{A}^{(t+r)}_{x^{+}_r}U_{t+r-1:t+r}\right],\\
  &\vert F^{SE}_{\bm{x}^-}\rangle\!\rangle = \sum_j \vert F^S_{\bm{x}^-\!,j}\rangle\!\rangle \otimes \vert F^E_{\bm{x}^-\!,j}\rangle\!\rangle =\left[\prod_{s=1}^{t-1}U_{s:s+1}\bar{A}^{(s)}_{x^{-}_s}\right] \vert \rho^{(0)}_{SE}\rangle\!\rangle,
\end{align}
implying the decomposition of $A^{(t)}_{x_t}$ on the non-orthogonal basis 
\begin{equation}
  \mathbb{B}^{(t)} = \left\{B_{\bm{x}^+\bm{x}^-} := \sum_{ij}\langle\!\langle F^E_{\bm{x}^+,i}\vert F^E_{\bm{x}^-,j}\rangle\!\rangle \vert F^S_{\bm{x}^-,j}\rangle\!\rangle\langle\!\langle F^S_{\bm{x}^+,i}\vert\right\},
\end{equation}
the LIST aims to establish a bijection between an integer, denoted as $\alpha$, and the concatenation of vectors $(\bm{x}^+,\bm{x}^-)$ through the adjustment of $\bm{x}^+$ and $\bm{x}^-$ such that $\mathbb{B}^{(t)}=\{B_0^{(t)}, B_1^{(t)},\dots, B_{d^4-1}^{(t)}\}$ is a linear independent basis set.  Consequently, LIST measures probabilities based on this constructed basis,
\begin{equation}\label{eq:inst_decomp_vec}
  \bm{p}_{x_t}^{(t)} = \begin{bmatrix}
    (\bm{b}_0^{(t)})^\dagger\\
    (\bm{b}_1^{(t)})^\dagger\\
    \vdots\\
    (\bm{b}_{d^4-1}^{(t)})^\dagger\\ 
  \end{bmatrix}\bm{a}^{(t)}_{x_t} = B^{(t)}\bm{a}^{(t)}_{x_t},
\end{equation}
where $\bm{a}^{(t)}_{x_t}$ and $\bm{b}_\alpha^{(t)}$ represent the vectorization of the $A^{(t)}_{x_t}$ and $B_\alpha^{(t)}$, respectively.  It should be noted that all instruments within an available set at time step $t$ share the same $B^{(t)}$. If $B^{(t)}$ is invertible, we can obtain instruments
\begin{gather}\label{eq:list_recover}
    \Xi^{(t)} =\left(B^{(t)}\right)^{-1} \Gamma^{(t)},
\end{gather}
where $\Xi^{(t)} = [\bm{a}^{(t)}_{0},\bm{a}^{(t)}_{1}, \dots, \bm{a}^{(t)}_{m_t-1}]$ and $\Gamma^{(t)} = [\bm{p}_{0}^{(t)},\bm{p}_{1}^{(t)},\dots,\bm{p}_{m_t-1}^{(t)}]$. Then, PTMs of instruments are recovered by devectorization of determined $\bm{a}^{(t)}_{x_t}$. 

The instruments are reconstructed by repeating this process for each time step. Then, we choose the maximum linear independent set of instruments $\mathbb{A}^{(t)}:=\{A^{(t)}_{j}\}$ at each time step to formulate the process tensor
\begin{align}
  \Upsilon_\mathcal{T} =\sum_{\bm{x}}p_{\bm{x}}\left(D^{(0)}_{x_0}\otimes\dots\otimes D^{(k-1)}_{x_{k-1}}\right),
\end{align}
where $\mathbb{D} : =\left\{D^{(t)}_{i}\right\}$ is the dual set of $\mathbb{A}^{(t)}$ such that $\mathrm{Tr}\left[\left(D^{(t)}_{i}\right)^\dagger A^{(t)}_{j}\right] = \delta_{ij}$. 

The tomography of the instrument set exhibits the gauge freedom up to a set of invertible matrices $\{B^{(t)}\}$ due to the inaccessible initial state and SE unitaries. Let $\left\{ \bm{q}^{(t)}_{x_{t}}\right\}$ denote the dual set of $\left\{\bm{p}^{(t)}_{x_{t}}\right\}$ that corresponds to $\left\{A^{(t)}_{x_{t}}\right\}$. From \eqref{eq:se_evo_prob_pt}, we can observe the probability invariance with respect to $\{B^{(t)}\}$,
\begin{align}
  p_{\bm{y}} =\sum_{\bm{x}}p_{\bm{x}}\prod_{t=0}^{k-1}\mathrm{Tr}\left[\left(D^{(t)}_{x_t}\right)^{\dagger}A^{(t)}_{y_t}\right]=\sum_{\bm{x}}p_{\bm{x}}\prod_{t=0}^{k-1}\left({\bm{q}}^{(t)}\right)^\dagger_{x_t}\left(B^{(t)}\right)^{-1} B^{(t)}\bm{p}^{(t)}_{y_t}.
\end{align}
It is impossible to distinguish the quantum operations up to $\{B^{(t)}\}$ based on the probability measurement, as one can obtain a set of instruments and process tensor without violations of measurement probabilities $p_{\bm{x}}$ for each given set of gauge matrices $\{B^{(t)}\}$. 

The gauge optimization is required to provide a reasonable set of gauge matrices for determining the tomographic result of the instrument set. Assuming that the quantum instruments are well-implemented and approximately close to the ideal instruments, the gauge matrix can be optimized by
\begin{equation}\label{eq:gauge_opt_obj_fn}
  B^{(t)} = \arg\min_X \sum_{t} \left\|X\Gamma^{(t)} - \Xi^{(t)}_{\mathrm{knowledge}}\right\|_F,
\end{equation}
where $\Xi^{(t)}_{\mathrm{knowledge}}$ represents the knowledge of instruments to the experimenter and $\|\cdot\|_F$ represent the Frobenius norm. Consequently, the tomographic result is
\begin{gather}
  \hat{\mathfrak{I}}=\left\{\hat{\mathcal{J}}, \hat{\mathcal{T}}\right\},\label{eq:result_list}\\
  \hat{\mathcal{J}}=\left\{\left\{A^{(t)}_{0},\dots,A^{(t)}_{m_t-1}\right\}\right\}_{t=0}^{k-1},\\
  \hat{\mathcal{T}}=\Upsilon_\mathcal{T}.
\end{gather}

There are several noteworthy points to consider. First, it can be observed from Eq.\eqref{eq:list_recover} and Eq.\eqref{eq:gauge_opt_obj_fn} that the LIST always determines instruments mathematically equal to the prior knowledge at the time step that instruments in the available set are linear independent and not overcomplete. In this scenario, the LIST degrades to the linear inversion PTT, and any imperfect implementation of instruments is represented by the process tensor. The LIST reveals the capability in detecting the disharmony of linear relationships when the instruments at a particular time step are not linear independent. Furthermore, the linear inverse method actually determines a self-consistent tomographic result of instruments and the process tensor, but they may not be physically implementable. These characteristics result from the absence of constraints in the gauge optimization. It is possible to impose constraints on each $B_\alpha^{(t)}$ and/or $A_{x_t}^{(t)}$ to be CPTNI\footnote{The constraints to the instruments are corresponding to the completely positive trace non-increasing assumption of instruments. This can be adjusted along with the instruments' assumptions (CPTP at intermediate time steps on NISQ devices, for example).}. Nevertheless, this will increase computational complexity. Instead, we optimize $B^{(t)}$ over the entire group of real and invertible matrices to strenuously fit the data. This is similar to the linear inverse GST (LGST) \cite{greenbaum2015Introduction} used in the Markovian setting.

Second, the objective function in \eqref{eq:gauge_opt_obj_fn} is not convex and may has nonunique global minima especially when the available set is not informationally complete. This indeterminacy is a common characteristic in quantum tomography, and it manifests in various scenarios such as QST, QPT, and GST. Besides the gauge freedom over $B^{(t)}$, there is another kind of gauge freedom that we cannot determine the initial state and actual SE unitaries by the LIST but a set of consistent ones, which is referred to the SPAM gauge freedom in LGST. In this case, fixing the gauge provides no additional information about the initial state and SE unitaries. Exploiting the process tensor from the reduced instrument set avoids the explicit discussion of the SPAM gauge freedom at each time step. Therefore, the LIST also derives the consequence that an initialization error of $\vert F^{SE}_{\bm{x}^-}\rangle\!\rangle$ cannot be distinguished from a faulty measurement $\langle\!\langle F^{SE}_{\bm{x}^+}\vert$ at each time step $t$, as described in GST \cite{greenbaum2015Introduction}. Furthermore, it is intractable to distinguish non-Markovian SE correlations before the intervention from non-Markovian SE correlations after.

Third, the LIST at each time step requires the an informationally complete basis spanning the space of instruments formed by combining instruments at other time steps, rather than that the system state and the measurement simultaneously and respectively form $d^2$-linear-independent sets which can be transformed into complete bases of $\mathcal{B}(\mathcal{H}_{S})$ the space of bounded linear operators on $\mathcal{H}_{S}$. This is because the environment carries the information by non-Markovian SE evolutions. However, it is challenging to confirm whether the entanglements before and after a time step are enough to convey the information, such that there exists a set of combinations of instruments at other time steps that have the capability to construct informationally complete basis for tomography at the time step, because the SE dynamics are inaccessible for experimenters. In other words, the non-Markovian effect may not be so severe as to have a high probability of satisfying the condition for composing a informationally complete basis. Therefore, we still recommend constructing complete basis on $\mathcal{H}_{S}$ before and after the time step, respectively. Note that this challenge becomes intractable when performing tomography at time steps closed to the beginning and end, particularly at time step $0$ and $k$, where the former or later instruments cannot generate a set of $\vert F^{SE}_{\bm{x}^-}\rangle\!\rangle$ or $\langle\!\langle F^{SE}_{\bm{x}^+}\vert$ that forms a $d^2$-linear-independent set. The proposed LIST does not address this issue. However, the tomographic results remain consistent with the measurement probabilities.

\subsection{Maximum Likelihood Estimation based IST}

A statistical framework for IST based on the maximum likelihood estimation (MLE-IST) is proposed. The MLE-IST is compatible with overcomplete data which may enhance the estimation. The tomographic results are guaranteed to be physically implementable by introducing constraints. Furthermore, experimenters may be interested in additional characteristics, such as the SE evolutions themselves, which require high flexibility of the model. As a result, the likelihood function of instrument set is derived as
\begin{align}
  l(\hat{\mathfrak{I}})=\sum_{\bm{x}}{\left(\tilde{p}_{\bm{x}}-\hat{p}_{\bm{x}}\right)^2}/{\sigma_{\bm{x}}^2},
\end{align}
where $\tilde{p}_{\bm{x}}$ denotes the measurement probability of performing instruments labeled by $\bm{x}$ obtained by the experiment, $\sigma_{\bm{x}}^2$ is the sampling variance of $\tilde{p}_{\bm{x}}$, and $\hat{p}_{\bm{x}}$ is the estimator of measurement probability which is modeled using parameters. The derivation of the likelihood function is given in the Appendix~\ref{appendix:likelihood}. The MLE-IST exhibits high flexibility estimating the instrument set with physical constraints based on the various assumptions, such as CPTNI for generality or CPTP on NISQ.

Based on the full definition of instrument set, as described in Eq.\eqref{eq:instrument_set_full}, each instrument, denoted as $\mathcal{A}_{x_t}^{(t)}$, is modeled by a real matrix $\hat{R}_{x_t}^{(t)}\in[-1,1]^{d^2\times d^2}$ as the PTM representation with CPTNI constraints. Specifically, the CP requires the Choi state of $\hat{R}_{x_t}^{(t)}$ to be positive semidefinite as 
\begin{align}
  \hat{\rho}_{x_t} = \frac{1}{d^2}\sum_{i,j=0}^{d^2-1}[\hat{R}_{x_t}^{(t)}]_{i,j} \left(P_j^T \otimes P_i\right) \succcurlyeq 0,
\end{align}
where $P_i$ represents the $i$-th Pauli matrix, $P_0 = I$. The TNI requires the first entry to be $0\le[\hat{R}_{x_t}^{(t)}]_{0,0} \le 1$. To parameterize SE unitaries, we introduce $\bm{\alpha}^{(t:t+1)} \in [-\pi,\pi]^{d^2 d^{\prime2}-1}$ corresponding to rotation angles of Pauli operators in $\mathbb{P} = \{P^{S}_i \otimes P^{E}_j|\forall i,j\}\setminus\{I^{S}\otimes I^{E}\}$. Then, the recovered unitary $\hat{V}_{t:t+1}(\bm{\alpha})$ is defined as the PTM of unitary $\exp\left(\iota (\bm{\alpha}^{(t:t+1)})^T\bm{\sigma}\right)$, where $\bm{\sigma}$ represents the vector of Pauli operators, $\iota^2 = -1$. The initial SE state $\vert\hat{\rho}^{(0)}_{SE}\rangle\!\rangle\in[-1,1]^{d^2d^{\prime2}\times 1}$ is modeled as a initial SE unitary $\hat{V}_{-1:0}(\bm{\alpha}^{(-1:0)})$ acting on the zero state $\vert\zeta_{SE}\rangle\!\rangle$. This model inherently ensures that the initial SE state is pure. For simplicity, we will use the notation $\hat{V}_{{t:t+1}}$ to indicate $\hat{V}_{{t:t+1}}(\bm{\alpha}^{(t:t+1)})$ in the following. Hence, the estimator of the probability is given by
\begin{align}
  \hat{p}_{\bm{x}}=\langle\!\langle 0_{SE}\rvert \bar{\hat{R}}^{(k-1)}_{x_{k-1}}\prod_{t=0}^{k-2} \left({\hat{V}_{t:t+1}\bar{\hat{R}}^{(t)}_{x_t}}\right) \hat{V}_{-1:0}\vert\zeta_{SE}\rangle\!\rangle.
\end{align}

Then, the optimization problem describing MLE-IST based on the full instrument set is given by 
\begin{align}\label{eq:mle_ist_opt_prob_full}
  \min&_{\substack{\lvert \hat{\rho}^{(0)}_{SE}\rangle\!\rangle,\hat{R}^{(t)}_{x_t}, \bm{\alpha}^{(t:t+1)},
  \forall x_t, t}}~l(\hat{\mathfrak{I}}),\\
  s.t.~& \hat{\rho}_{x_t} = \frac{1}{d^2}\sum_{i,j=0}^{d^2-1}[\hat{R}_{x_t}^{(t)}]_{i,j}\left(P_j^T \otimes P_i\right) \succcurlyeq 0, \forall x_t, \label{eq:full_def_cp_costraint}\\
  &0\le[\hat{R}_{x_t}^{(t)}]_{1,1} \le 1, \forall x_t,\label{eq:full_def_tni_costraint}\\
  &-1\le[\hat{R}_{x_t}^{(t)}]_{i,j} \le 1, \forall x_t,i,j, \label{eq:full_def_ptm_val_constraint}\\
  &-\pi \le [\bm{\alpha}^{(t:t+1)}]_i \le \pi, \forall t, i,\label{eq:full_def_upval_constraint}
\end{align}
where \eqref{eq:full_def_cp_costraint} and \eqref{eq:full_def_tni_costraint} constrain the instruments to be CP and TNI, respectively, \eqref{eq:full_def_ptm_val_constraint} defines the range of PTM entries, and \eqref{eq:full_def_upval_constraint} limits the range of parameters of SE unitaries and the initial state. Consequently, the MLE-IST estimates the instrument set as 
\begin{gather}\label{eq:result_mleist_full}
  \hat{\mathfrak{I}}:= \left\{\hat{\mathcal{J}}, \hat{\mathcal{U}},\hat{V}_{-1:0}\vert\zeta_{SE}\rangle\!\rangle\}\right\}\\
  \hat{\mathcal{J}}= \left\{\left\{\hat{R}^{(t)}_{0},\dots, \hat{R}^{(t)}_{m_t-1}\right\}\right\}_{t=0}^{k-1},\\
  \hat{U} = \left\{\hat{V}_{t:t+1}\right\}_{t=0}^{k-2},
\end{gather}
{with $\sum_{t=0}^{k-1}m_t d^4 + k(d^2d^{\prime2}-1)$ parameters. Notably, the number of parameters can be further reduced due to the fact that the dimensions of SE correlations in the recent devices are limited. For example, dimensions of the environment can be set identically to the system for recent NISQ devices, i.e. $d^\prime = d$. Therefore, the totol parameters are $\sum_{t=0}^{k-1}m_t d^4 + k(d^4-1)$, which is polynomial to the time step $k$. See Section~\ref{sec:result} for detail.}

The models of instruments and the SE unitaries offer flexibility, allowing for manipulation based on assumptions and characteristics of the instrument set of interest. For example, instrument constraints can be substituted with CPTP constraints for quantum gates on NISQ devices, while measurement instruments can be treated as vectors. Additionally, the SE unitary can be modeled as a CPTP real orthonormal matrix.

The optimization problem is evidently non-convex and may have multiple global optima due to the multiplications of variable matrices in $\hat{p}_{\bm{x}}$ leading to $k$-order polynomial with $k$-order exponential parameters. Hence, an appropriate initialization of parameters is of considerable importance in improving the optimization process. We suggest initializing the MLE-IST by the result of LIST with identity initialization of $\hat{V}_t$. If the LIST yields nonphysical outcomes, results of regular MLE-GST under the Markovian assumption are recommended to initialize MLE-IST. 

Additionally, the MLE-IST is capable of working with the reduced instrument set, which requires $\mathcal{O}(d^{4k})$ parameters. It is computationally intractable to solve the problem with the massive parameters {without further compressive considerations}. Therefore, the MLE-IST framework with the reduced instrument set is derived in Appendix~\ref{appendix:reduced_mle}, but its implementation for simulations and experiments is not pursued.

\section{Performing IST on NISQ Devices}\label{sec:nisq_ist}
A typical NISQ device executes a given quantum circuit consisting of a sequence of quantum gates at intermediate time steps and measurements at the end, where quantum gates are unitaries and CPTP operations in the ideal and noisy settings, respectively, and each measurement is a 2-valued POVM on a single qubit. Hence, the available set at time step $t$ consists of a set of CPTP operations and a set of measurements,
\begin{align}\label{eq:nisq_inst}
  \mathcal{J}^{(t)} := \left\{\left\{\mathcal{A}^{(t)}_{x_t}\right\}, \left\{\mathcal{M}_{x_t}^{(t)}\right\}\right\}.
\end{align}
Then, the measurement probability for a $k$-time step non-Markovian quantum circuit is given by
\begin{align}\label{eq:nisq_exp}
  p_{\bm{x}} = \langle\!\langle \bar{M}^{(k-1)}_{x_{k-1}}\vert\prod_{t=0}^{k-2} {U_{t:t+1}\bar{A}^{(t)}_{x_t}} \vert\rho^{(0)}_{SE}\rangle\!\rangle,
\end{align}
where $\langle\!\langle \bar{M}^{(t)}_{x_{t}}\vert = \langle\!\langle {M}^{(t)}_{x_{t}}\vert \otimes \langle\!\langle 0_{E}\vert$.
Associating with Eq.\eqref{eq:nisq_exp} and Eq.\eqref{eq:se_evo_prob_tr}, the measurement $\langle\!\langle M_{x_t}^{(t)}\rvert$ corresponds to the first row of the PTM of a CP and trace decreasing (TD) instrument at the end, with all other entries remaining $0$. Therefore, LIST can be performed directly by representing measurements as regular instruments at the last time step and extracting the first row as the result, when the non-Markovian correlation is sufficient to construct informationally complete basis by adjusting the instruments before the time step. However, it becomes intractable in other cases. This limitation arises from the requirement that the measurement must be the final instrument in the circuit. Thus, in the LIST strategy, each time step is treated as the last time step.

Based on the observation that the PTM matrix of a (CPTD) measurement is always linear independent from the CPTP maps, the LIST can be implemented by treating CPTP maps and measurements separately. The first row of CPTP maps is omitted in the vectorization in Eq.~\eqref{eq:inst_decomp_vec}, resulting in the requirement of $d^2(d^2-1)$ measured probabilities per CPTP map and $d^2(d^2-1)\times d^2(d^2-1)$ dimensional gauge matrix. Then, the tomography of measurements is conducted by $d^2$ measured probabilities per measurement and $d^2\times d^2$ dimensional gauge matrix. Probability data and gauge matrix for CPTP instruments and measurements are measured and optimized in separate subroutines independently, while the remaining steps follow the ordinary LIST approach.

As for MLE-IST, the model is simplified to enhance the efficiency. Each measurement can be modeled by a $d^2$-dimensional real row vector $\langle\!\langle \hat{E}_{x_t}\vert\in[-1, 1]^{1\times d^2}$ with positive constraints, i.e., both the matrix $\hat{E}_{x_t}$ and $I-\hat{E}_{x_t}$ are positive semidefinite. Each intermediate instrument can be modeled by $d^2\times (d^2-1)$ parameters with CP constraint, while the TP constraint implies that the first row of $\hat{R}_{x_t}^{(t)}$ is $[1,0,0,\dots,0]$. Then, the estimator of probability is given by
\begin{align}\label{eq:nisq_estimator}
  \hat{p}_{\bm{x}} = \langle\!\langle \bar{\hat{E}}^{(k-1)}_{x_{k-1}}\vert\prod_{t=0}^{k-2} {\hat{V}_{t:t+1}\bar{\hat{R}}^{(t)}_{x_t}} \vert\hat{\rho}^{(0)}_{SE} \rangle\!\rangle.
\end{align}

Consequently, the optimization problem for MLE-IST on NISQ devices can be described as
\begin{align}
  \min&_{\substack{\lvert \hat{\rho}^{(0)}_{SE}\rangle\!\rangle, \langle\!\langle \hat{E}_{x_t}^{(t)} \vert,\hat{R}^{(t)}_{x_t},\hat{V}_{t:t+1}, 
  \forall x_t, t}}~l(\hat{\mathcal{I}}),\\
  s.t.~& \eqref{eq:full_def_cp_costraint}, \eqref{eq:full_def_ptm_val_constraint}, \eqref{eq:full_def_upval_constraint}\notag\\
  &[\hat{R}_{x_t}^{(t)}]_{0,i} = \delta_{0,i}, \forall x_t, i,\label{eq:cptp_constraint}\\
  &\hat{E}^{(t)}_{x_t}= \frac{1}{\sqrt{d}} \sum_{i=0}^{d^2-1} \langle\!\langle \hat{E}_{x_t}^{(t)} \vert i\rangle\!\rangle P_i \succcurlyeq 0,~ \forall t,\label{eq:mea_positive}\\
  &I-\hat{E}_{x_t} \succcurlyeq 0,~ \forall t,\label{eq:comp_mea_positive}
\end{align}
where Eq.~\eqref{eq:cptp_constraint} is the CPTP constraint, Eq.~\eqref{eq:mea_positive} and Eq.~\eqref{eq:comp_mea_positive} are positive constraints of measurements.

\section{Experimental Results}\label{sec:result}

In this section, we perform numerical simulations and real-chip experiments on $4$-time-step single-qubit systems to demonstrate the effectiveness of IST. All simulations and experiments are conducted under NISQ situation, as described in Section~\ref{sec:nisq_ist}. Particularly, the knowledge and perfect implementation of an intermediate instrument are mathematically a unitary operation without any ancilla qubits. However, the imperfect (real-chip) implementation is a noisy CPTP operator. Based on the real quantum device provided by IBM Quantum Experience (QX) \cite{ibmquantum2021Ibmq_belem,ibmquantum2021Ibm_perth} and Origin Quantum (OriginQ) \cite{originquantum2021Origin_wuyuan_d4,originquantum2020Origin_wuyuan_d5}, utilized instruments and their knowledge are listed in the Table~\ref{table:utilized_inst}, where $\{\mathcal{A}_0,\dots,\mathcal{A}_{11},\mathcal{M}\}$ are native to IBM QX devices, and $\{\mathcal{A}_0,\dots,\mathcal{A}_{7},\mathcal{A}_{12},\dots,\mathcal{A}_{14},\mathcal{M}\}$ are physically implementable to OriginQ devices.

\begin{table}[h!]
  \centering
  \caption{Notations and knowledge of utilized instruments. The knowledge of instruments are represented by quantum gates and projected measurements.}
  \begin{tabular}{c| c c c c} 
   notation & $\mathcal{A}_0$ & $\mathcal{A}_1$ & $\mathcal{A}_2$ & $\mathcal{A}_3$  \\ [0.5ex] 
   knowledge & $I$ & $R_Z\!\left(\frac{\pi}{2}\right)$ & $R_Z\!\left(\frac{\pi}{4}\right)$ & $R_Z\!\left(\frac{\pi}{6}\right)$ \\ [0.5ex] 
   \hline
   notation & $\mathcal{A}_4$ & $\mathcal{A}_5$ & $\mathcal{A}_6$ & $\mathcal{A}_7$\\[0.5ex] 
   knowledge & $X$ & $XR_Z\!\left(\frac{\pi}{2}\right)$ & $XR_Z\!\left(\frac{\pi}{4}\right)$ & $XR_Z\!\left(\frac{\pi}{6}\right)$\\[0.5ex] 
   \hline
   notation & $\mathcal{A}_8$ & $\mathcal{A}_9$ & $\mathcal{A}_{10}$ & $\mathcal{A}_{11}$ \\ [0.5ex]
   knowledge & $\sqrt{X}$ & $\sqrt{X}R_Z\!\left(\frac{\pi}{2}\right)$ & $\sqrt{X}R_Z\!\left(\frac{\pi}{4}\right)$ & $\sqrt{X}R_Z\!\left(\frac{\pi}{6}\right)$\\[0.5ex] 
   \hline
   notation & $\mathcal{A}_{12}$ & $\mathcal{A}_{13}$ & $\mathcal{A}_{14}$ & $\mathcal{M}$ \\[0.5ex] 
   knowledge& $H$ & $HR_Z\!\left(\frac{\pi}{2}\right)$ & $HR_Z\!\left(\frac{\pi}{4}\right)$ & $\vert 0 \rangle\langle 0\vert$ \\ [0.5ex] 
  \end{tabular}
  \label{table:utilized_inst}
  \end{table}

The criterion to benchmark the effectiveness of tomography is the square error of probabilities (SEP)
\begin{align}\label{eq:square_error}
  S(\hat{\mathfrak{I}})=\sum_{\bm{x}}{\left(\tilde{p}_{\bm{x}}-\hat{p}_{\bm{x}}\right)^2},
\end{align}
where $\hat{p}_{\bm{x}}$ represents the recovered probability of $\bm{x}$ computed by tomographic result $\hat{\mathfrak{I}}$, $\tilde{p}_{\bm{x}}$ is the measurement probability, and the range of $x$ is combinations of all instruments. Although using true probabilities instead of measured probabilities in \eqref{eq:square_error} is intrinsically better to benchmark the effectiveness, the experimenter cannot directly access true probabilities. Nonetheless, this criterion is still appropriate as it measures the consistency of the tomographic result to the measured probabilities. 

To showcase the advancement of the proposed framework, we implement process tensor tomography (PTT) and maximum likelihood estimation based gate set tomography (MLE-GST) for transversal comparison with LIST and MLE-IST, respectively. The technical details for the methods to be compared are summarized as follows:
\begin{itemize}
  \item \textbf{PTT.} Constructing process tensor by the dual set of specified instruments and corresponding probabilities;
  \item \textbf{MLE-GST.} MLE-GST exploits maximum likelihood estimation with physical constraints to estimate instruments that are physically implementable. Note that quantum gates in different time steps are independently estimated;
  \item \textbf{MLE-GST-S.} A simplified version of MLE-GST where quantum gates with the same name in different time steps are assumed to be identical.
\end{itemize}



\subsection{Numerical Simulations}
Numerical simulation circuits are constructed as depicted in Figure~\ref{fig:qsp} using the Pennylane package in Python. {Dimensions of environment are set identically as the system for facilitating the demonstration of the effectiveness of our proposed methods.} The initial SE state is generated by applying the initial SE unitary on the zero state as $\rho_{SE}^{(0)} = \mathcal{U}_{-1:0}\left(\vert 0_{SE}\rangle\langle 0_{SE} \vert\right)$. For all time steps, SE unitaries are selected from
\begin{align}
  U_0 &:= R_{ZZ}(0.2)R_{YY}(0.2)R_{XX}(0.2),\\
  U_1 &:= R_{ZZ}(0.2)R_{XY}(0.2),\\
  U_2 &:=  R_{IX}(0.2)R_{YY}(0.2)R_{XX}(0.2)R_{XI}(0.2),
\end{align}
where $R_{P^SP^E}(\theta):=\exp(\frac{-\iota \theta P^SP^E}{2})$. Note that $U_0$ preserves the zero state, indicating the absence of SE correlations in the initial SE state when $\mathcal{U}_{-1:0}$ is set as $U_0$. In the case of imperfect implementations, a depolarizing channel and an amplitude damping channel are applied after each CPTP instrument, with noise parameters increasing linearly with the time step as $0.05(t+1)$.

\subsubsection*{Result of LIST simulations}

Simulations for LIST are performed with available sets
\begin{equation}\label{eq:overcomplete_instruments}
  \mathcal{J}^{(t)}_{oc} := \left\{\begin{aligned}&\left\{\mathcal{A}_0, ..., \mathcal{A}_{11}, \mathcal{M}\right\}, & t\ne k-1,\\
    &\left\{\mathcal{M}\right\}, & t=k-1,\end{aligned}\right.\\
\end{equation}
where CPTP instruments in $\mathcal{J}^{(t)}_{oc}\vert_{t<k-1}$ are overcomplete to span the space of single-qubit unitary. Moreover, all instruments in $\mathcal{J}^{(t)}_{oc}$ are perfectly implemented as knowledge except for $\mathcal{A}_3$, $\mathcal{A}_7$ and $\mathcal{A}_{11}$, where $R_{Z}\!\left(\pi/6\right)$ are biased to $R_{Z}\!\left(\pi/5\right)$. The SE unitaries are uniformly set as $U_0$, resulting in an incomplete decomposition basis for CPTP operators in the first time step, while being complete for all subsequent intermediate time steps. 

\begin{figure*}[t]
  \centering
	\includegraphics[width=0.95\textwidth]{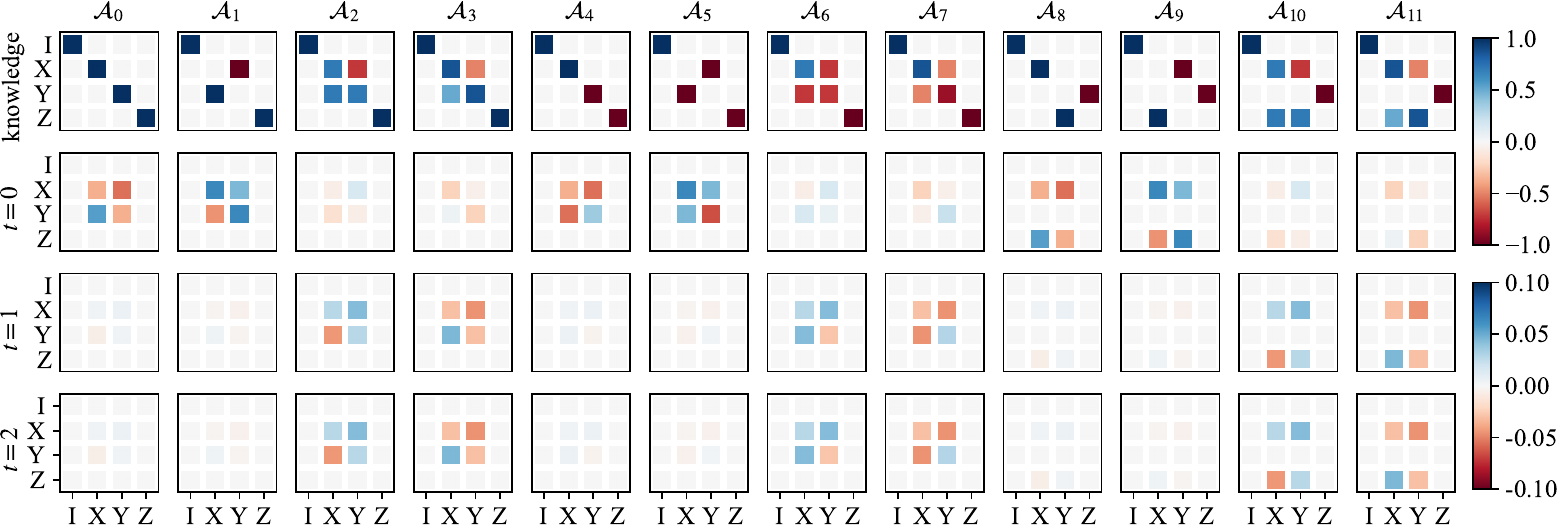}
  \caption{\label{fig:oc_list_diff} Differences between the knowledge and tomographic results of LIST with overcomplete instruments. Specifically, biases in the rotations on the Pauli Z-axis are present in $Z$ of $\mathcal{A}_3$, $\mathcal{A}_7$ and $\mathcal{A}_{11}$. It is important to note the scale of the color bar for accurate interpretation. The measurements and process tensor results are not shown here, as the former's tomographic outcome always matches the knowledge because there is only one single measurement, and the latter's detail is not significant enough to be depicted.}
\end{figure*}

Differences between the knowledge and tomographic results of CPTP maps are illustrated in Figure~\ref{fig:oc_list_diff} with SEP of $8.3\times 10^{-17}$. From the results, the catastrophic output of instruments at time step $0$ are revealed, due to the insufficiency of initial SE correlations for constructing a complete decomposition basis spanning the space of unitary. However, the LIST successfully estimates the CPTP maps at time steps $1$ and $2$ as expected. Remarkably, differences between the knowledge and implementations of instruments $\mathcal{A}_3$, $\mathcal{A}_7$ and $\mathcal{A}_{11}$ are detected. However, this disharmony not only impacts the tomographic results of $\mathcal{A}_3$, $\mathcal{A}_7$ and $\mathcal{A}_{11}$, since the LIST cannot exactly distinguish which instruments are incorrectly implemented. Furthermore, the non-unique global optima result in outcomes that are consistent with probabilities but deviate from the corresponding prior PTMs.

\begin{figure*}[t]
  \centering
	\subfigure[]{
		\begin{minipage}[h]{0.45\textwidth}
		\centering
		\includegraphics[width=\textwidth]{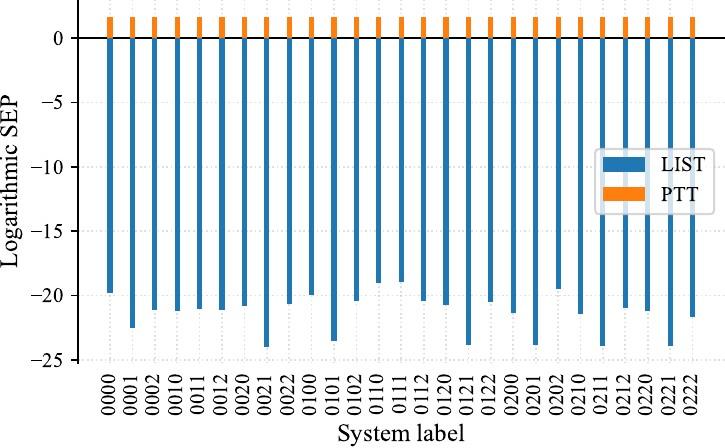}
		\end{minipage}
	}
  \subfigure[]{
		\begin{minipage}[h]{0.45\textwidth}
		\centering
		\includegraphics[width=\textwidth]{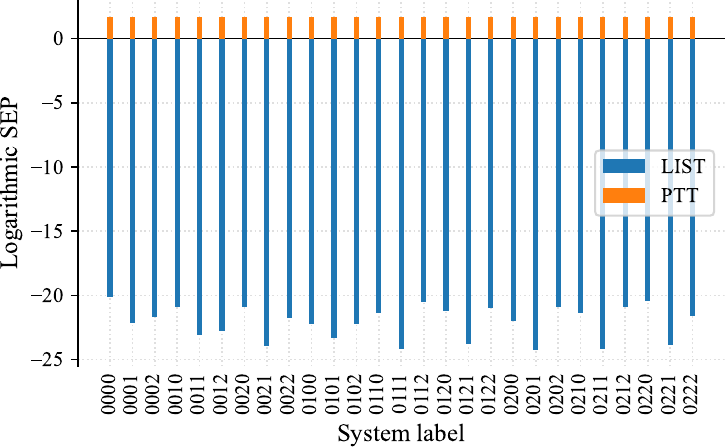}
		\end{minipage}
	}
  \caption{\label{fig:list_se_comp} The logarithmic SEP of LIST and PTT with respect to non-Markovian systems. (a) Instruments are perfectly implemented as knowledge. (b) Instruments are imperfectly implemented with specified quantum noise. Instruments $\mathcal{A}_3$, $\mathcal{A}_7$ and $\mathcal{A}_{11}$ in both (a) and (b) are implemented with biased rotations on Pauli $Z$.}
\end{figure*}

Then, we implement LIST and PTT for a batch of non-Markovian systems, with $\mathcal{J}^{(t)}_{oc}$ defined as in \eqref{eq:overcomplete_instruments}. Rotations on Pauli $Z$ in $\mathcal{A}_3$, $\mathcal{A}_7$ and $\mathcal{A}_{11}$ are still biased. Each non-Markovian system is labeled by a string of integers $u_{0}u_{1}\dots u_{k}$, representing the SE unitaries $\mathcal{U}_{-1:0},\dots,\mathcal{U}_{k-1:k}$ as $U_{u_{0}},\dots,U_{u_{k}}$. For PTT, we select $\mathcal{J}^{(t)}_{oc}\setminus \{\mathcal{A}_2, \mathcal{A}_6, \mathcal{A}_{10}\}$ for time steps $t<k-1$ to construct the process tensor. Figure~\ref{fig:list_se_comp} illustrates the SEP of simulations with perfect and imperfect instrument implementations. The LIST achieved significantly lower SEP compared to PTT, with average order reductions of -23.03 and -23.77 in perfect and imperfect situations, respectively, demonstrating self-consistency with the instruments that exist disharmony of linear relationships.

\subsubsection*{Result of MLE-IST simulations}

Simulations for MLE-IST are conducted with complete instruments. Based on native quantum gates on quantum devices of IBM QX, available sets are constructed by 
\begin{equation}\label{eq:ibm_complete_instruments}
  \mathcal{J}^{(t)}_{\mathrm{ibm}} := \left\{\begin{aligned}&\left\{\mathcal{A}_0,\mathcal{A}_1,\mathcal{A}_2,\mathcal{A}_4,\mathcal{A}_5,\mathcal{A}_6, \mathcal{A}_8, \mathcal{A}_9, \mathcal{A}_{10}, \mathcal{M}\right\}, & t\ne k-1,\\
    &\left\{\mathcal{M}\right\}, & t=k-1,\end{aligned}\right.\\
\end{equation}
where CPTP instruments in $\mathcal{J}^{(t)}_{oc}\vert_{t<k-1}$ are complete to span the space of single-qubit unitary. 

\begin{figure*}[t]
  \centering
	\includegraphics[width=0.95\textwidth]{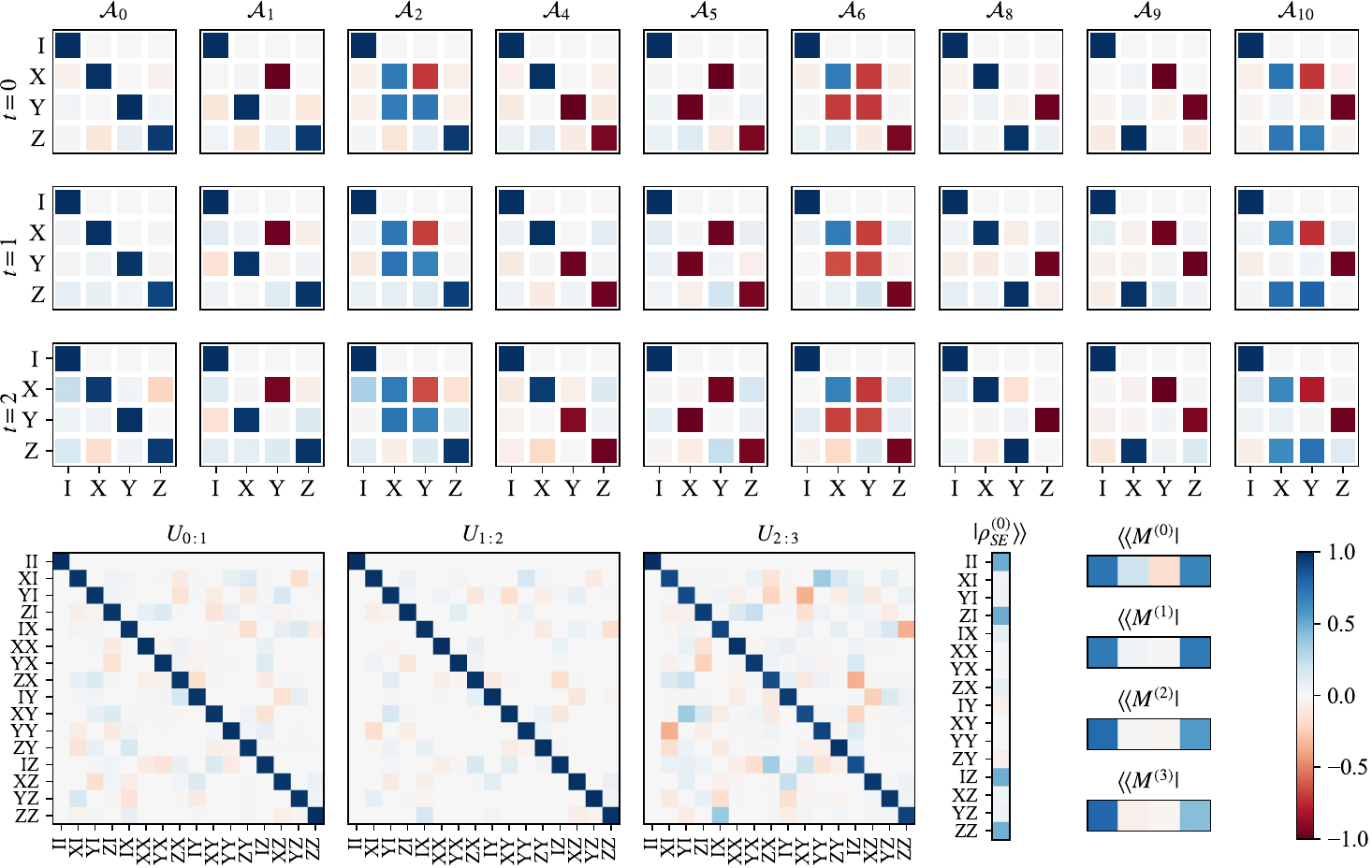}
  \caption{\label{fig:imperfect_inst_set} Tomographic result of MLE-IST with imperfect implementations of complete instruments. Color cells in each subfigure represent values from $-1$ to $1$ as shown in the color bar at the bottom right. The first three rows from top of the figure demonstrate PTMs of instruments (c.f. Table~\ref{table:utilized_inst} and \eqref{eq:ibm_complete_instruments}) at time step $t=0,1,2$. Subfigures labeled by $U_{t:t+1}$ represent the PTMs of SE unitaries $\mathcal{U}_{t:t+1}$, $t=0,1,2$. Subfigures labeled by $\vert\rho_{SE}^{(0)}\rangle\!\rangle$ and $\langle\!\langle M^{(t)}\vert$ represent superoperators of initial SE state and measurements at time step $t=0,1,2,3$.}
\end{figure*}

Specifying SE unitaries identically as $U_0$, the tomographic result of MLE-IST with imperfect instrument implementations is presented in Figure~\ref{fig:imperfect_inst_set}, with SEP of $8.0\times 10^{-8}$, which shows that the IST methods effectively reconstruct the instrument set. However, there are non-ignorable differences between the setups and the results of SE unitaries, initial SE state, and measurements at the start and the end time steps. This may arise from the non-uniqueness of the result, exacerbated by the insufficient SE correlations to construct a complete decomposition basis. In such cases, multiple unitaries and initial states can fit the probability data, leading to potential deviations from the experimenter's expectations while remaining faithful to the probability data.

\begin{figure*}[t]
  \centering
	\subfigure[]{
		\begin{minipage}[h]{0.46\textwidth}
		\centering
		\includegraphics[width=\textwidth]{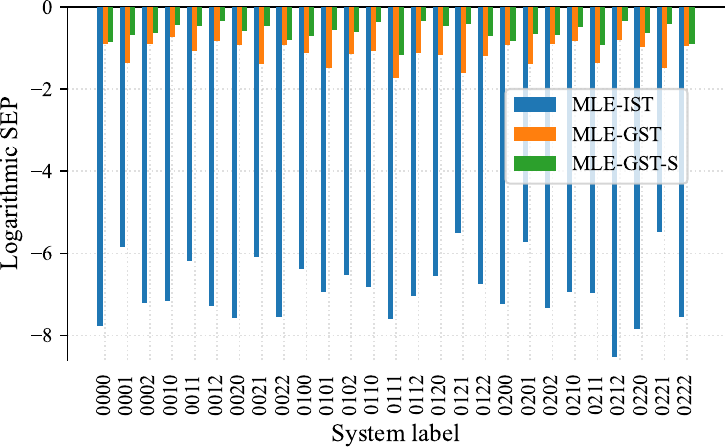}
		\end{minipage}
	}
  \subfigure[]{
		\begin{minipage}[h]{0.46\textwidth}
		\centering
		\includegraphics[width=\textwidth]{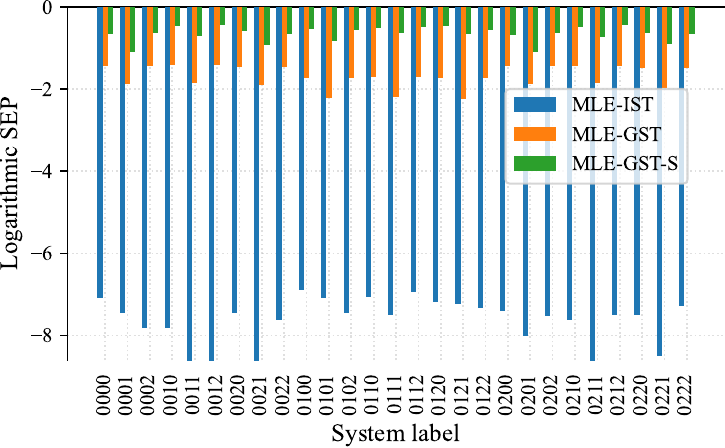}
		\end{minipage}
	}
  \caption{\label{fig:mle_ist_se_comp} The logarithmic SEP of MLE-IST, MLE-GST, and MLE-GST-S with respect to non-Markovian systems. (a) Instruments are perfectly implemented as knowledge. (b) Instruments are imperfectly implemented with specified quantum noise.}
\end{figure*}

Then, we implement MLE-IST, MLE-GST, and MLE-GST-S for a batch of non-Markovian systems with $\mathcal{J}^{(t)}_{\mathrm{ibm}}$. The SEP of simulations with perfect and imperfect instrument implementations is depicted in Figure~\ref{fig:mle_ist_se_comp}. Compared to MLE-GST and MLE-GST-S, MLE-IST shows significant improvements in SEP with reduction by orders of -5.78 (perfect) and -6.21 (imperfect) for MLE-GST, and by orders of -6.30 (perfect) and -7.24 (imperfect) for MLE-GST-S. The notably low SEP of MLE-IST underscores its essential capability to characterize non-Markovian correlations.

\subsection{Real-Chip Experiments}

\begin{figure*}[t]
  \centering
	\subfigure[]{
		\begin{minipage}[h]{0.45\textwidth}
		\centering
		\includegraphics[width=\textwidth]{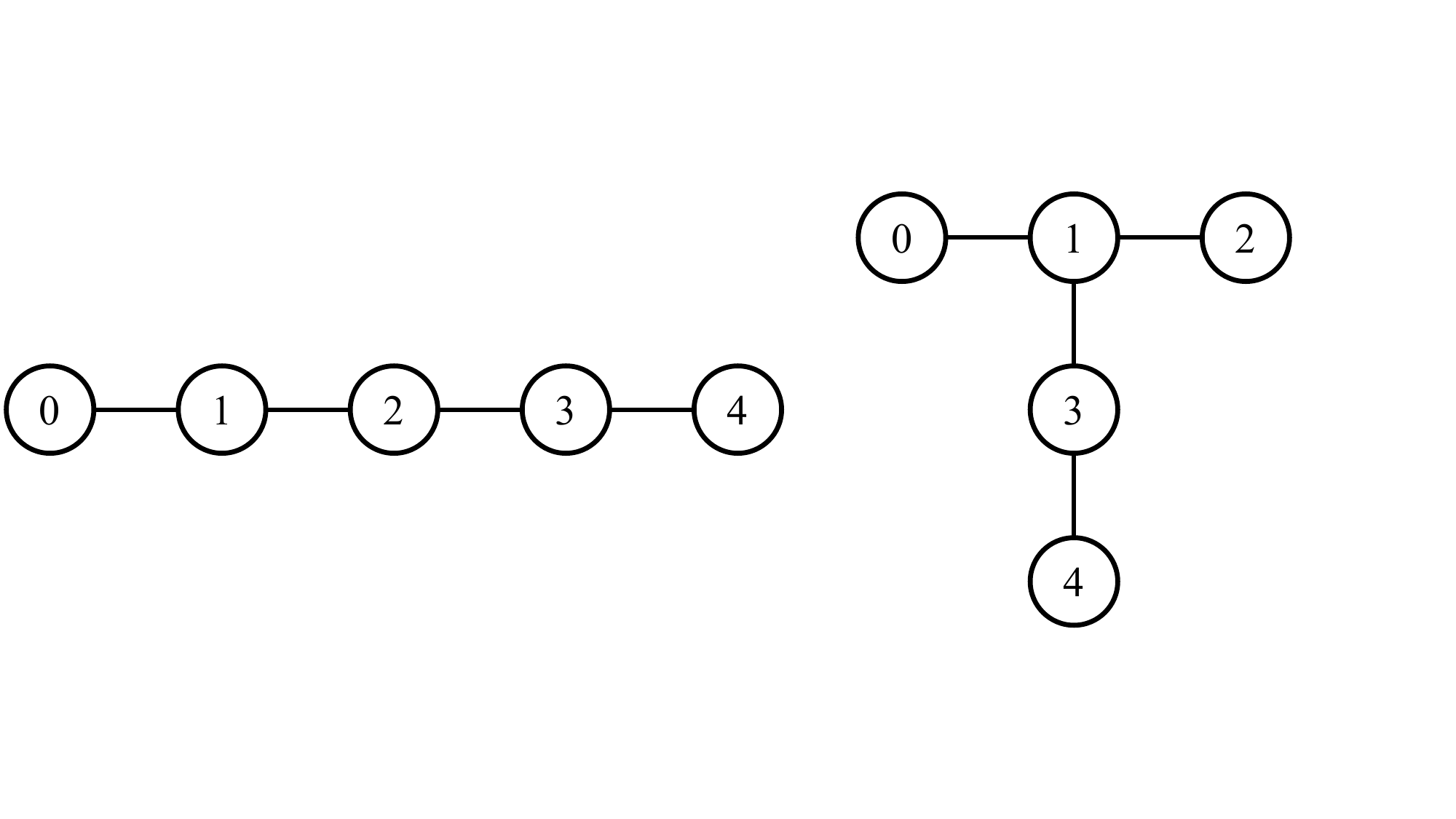}
		\end{minipage}
	}
  \subfigure[]{
		\begin{minipage}[h]{0.25\textwidth}
		\centering
		\includegraphics[width=\textwidth]{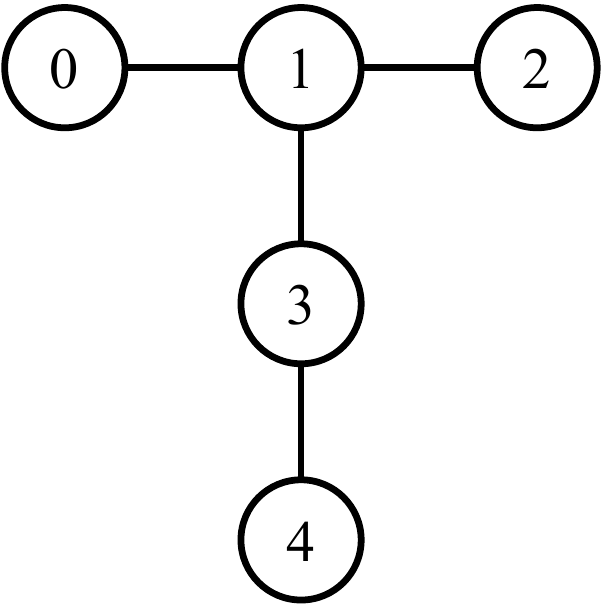}
		\end{minipage}
	}
  \caption{\label{fig:coupling} Coupling topology of quantum devices. (a) Coupling graph of {origin\_wuyuan\_d4} and {origin\_wuyuan\_d5 }. (b) Coupling graph of {ibmq\_belem} and {ibm\_perth}}
\end{figure*}

We conducted experiments on real quantum devices to implement the IST and comparative methods. \respred{Specifically, we implement LIST and PTT on OriginQ superconductive devices labeled by {origin\_wuyuan\_d4} \cite{originquantum2021Origin_wuyuan_d4} and {origin\_wuyuan\_d5 }\cite{originquantum2020Origin_wuyuan_d5} with linear coupling topology as shown in Figure~\ref{fig:coupling}(a). The MLE-IST, MLE-GST, and MLE-GST-S are implemented on IBM QX devices labeled by {ibmq\_belem} \cite{ibmquantum2021Ibmq_belem} and {ibm\_perth} \cite{ibmquantum2021Ibm_perth} with T-type coupling topology as shown in Figure~\ref{fig:coupling}(b).} 

\respred{During these experiments, a time step of intermediate instrument intervention consists of $T$ slots of quantum gates specified by the experimenter. At each time step, an instrument consisting of a sequence of quantum gates is implemented first. Then, a list of identity gates are performed after the instrument to guarantee the number of slots in a time step to be $T$.}

\subsubsection*{Result of LIST experiment}

The LIST is implemented using incomplete and linear dependent instruments on OriginQ devices labeled by {origin\_wuyuan\_d4} and {origin\_wuyuan\_d5}.
Available sets are subsets of $\mathcal{J}^{(t)}_{oc}$ that 
\begin{equation}\label{eq:incomplete_instruments}
  \mathcal{J}^{(t)}_{ic} := \left\{\begin{aligned}&\left\{\mathcal{A}_0, ..., \mathcal{A}_{7}, \mathcal{M}\right\}, & t\ne k-1,\\
    &\left\{\mathcal{M}\right\}, & t=k-1,\end{aligned}\right.\\
\end{equation}
where CPTP instruments in $\mathcal{J}^{(t)}_{ic}\vert_{t<k-1}$ are linear dependent. For PTT, we select the maximum linear independent set of instrument $\mathcal{J}^{(t)}_{ic}\setminus \{\mathcal{A}_3, \mathcal{A}_7\}$ for time steps $t<k-1$ to construct the process tensor. Each circuit is measured by independently sampling 5000 times.

\begin{figure*}[t]
  \centering
	\subfigure[]{
		\begin{minipage}[h]{0.46\textwidth}
		\centering
		\includegraphics[width=\textwidth]{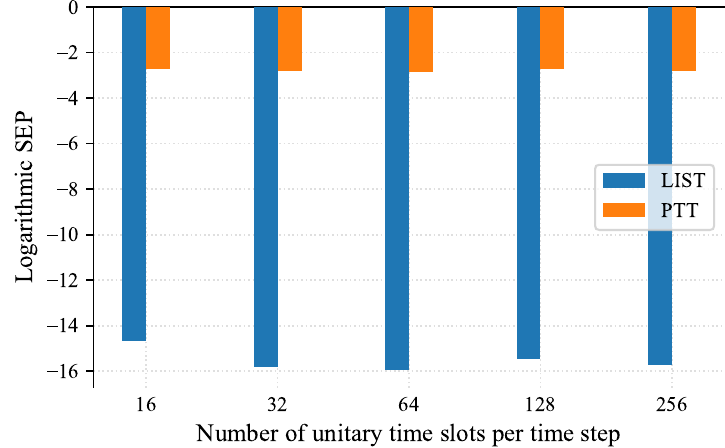}
		\end{minipage}
	}
  \subfigure{
		\begin{minipage}[h]{0.46\textwidth}
		\centering
		\includegraphics[width=\textwidth]{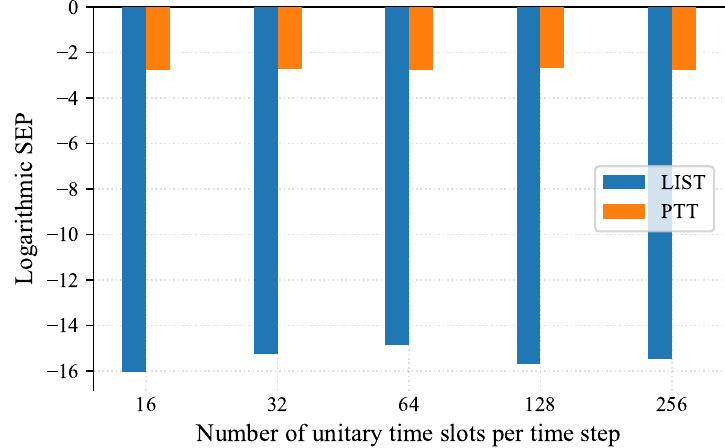}
		\end{minipage}
	}
  \caption{\label{fig:list_origin_se_comp} The logarithmic SEP of LIST and PTT with respect to the number of time slots on quantum computer (a) {origin\_wuyuan\_d4} and (b) {origin\_wuyuan\_d5}.}
\end{figure*}

The SEP results of LIST experiments are exhibited in Figure~\ref{fig:list_origin_se_comp}. From these results, the LIST maintains a low SEP close to 0, consistent with the simulation results, while sacrificing physical plausibility. Compared to PTT, LIST achieves an average improvement of SEP reduction by orders of -12.76 and -12.74 on {origin\_wuyuan\_d4} and {origin\_wuyuan\_d5}, respectively, highlighting the effectiveness of our proposed method. The significant gap between the SEPs of LIST and PTT indicate that the disharmony of linear relationship of instruments on the real quantum devices should be detected when performing tomography.

\subsubsection*{Result of MLE-IST experiment}

We implement the MLE-IST, MLE-GST, and MLE-GST-S on both IBM QX devices labeled by {ibmq\_belem} and {ibm\_perth}, exploiting available sets $\mathcal{J}^{(t)}_{\mathrm{ibm}}$ defined in \eqref{eq:ibm_complete_instruments}. Each circuit is measured by independently sampling 10000 times.

\begin{figure*}[t]
  \centering
	\subfigure[]{
		\begin{minipage}[h]{0.46\textwidth}
		\centering
		\includegraphics[width=\textwidth]{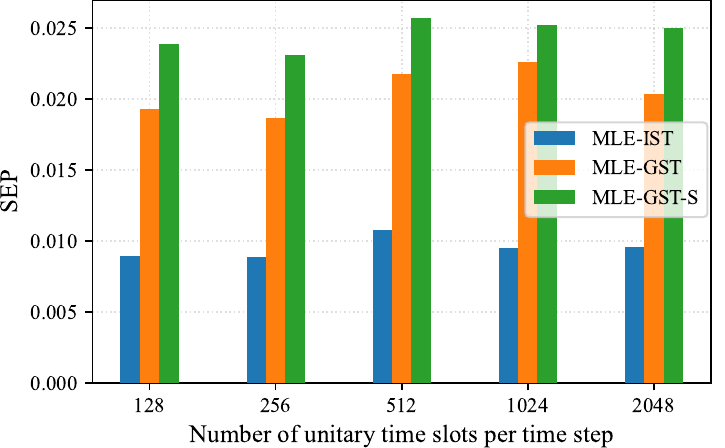}
		\end{minipage}
	}
  \subfigure[]{
		\begin{minipage}[h]{0.46\textwidth}
		\centering
		\includegraphics[width=\textwidth]{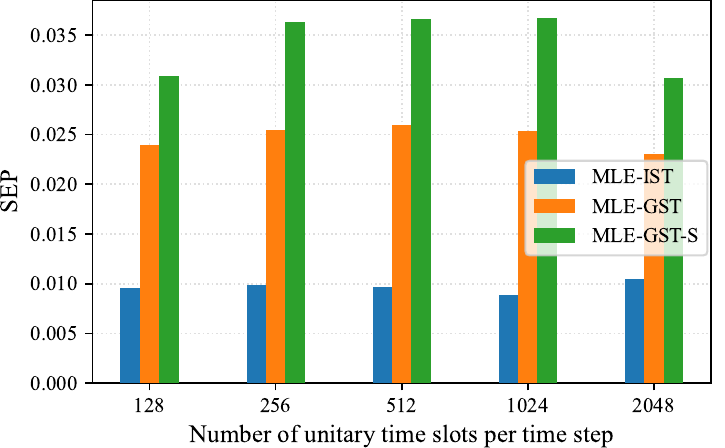}
		\end{minipage}
	}
  \caption{\label{fig:mle_rd_se_comp} The SEP of MLE-IST, MLE-GST and MLE-GST-S with respect to the number of time slots on quantum computer: (a) {ibmq\_belem} and (b) {ibm\_perth}.}
\end{figure*}

As shown in Figure~\ref{fig:mle_rd_se_comp}, the MLE-IST achieves average improvements on SEP of 53.55\% and 60.84\% on ibm\_belem and ibm\_perth, respectively, compared to MLE-GST. This verifies the applicability of the MLE-IST framework on real NISQ devices. The mitigated SEP compared with other tomographic schemes exhibits the necessity of MLE-IST that simultaneously estimates accessible instruments and inaccessible SE correlations. {Remarkably, these results indicate that a polynomial number of parameters with respect to the Markovian order are sufficient to characterize non-Markovian quantum noise in current NISQ devices, which certifies the appropriateness of the setting of the environment dimensions.}

\section{Discussion}\label{sec:discussion}

In this paper, we proposed a framework for instrument set tomography (IST) for quantum gate set tomography under the non-Markovian situation. Based on the quantum stochastic process operationally representing the non-Markovian quantum correlation and evolution, the instrument set is defined in full and reduced formation. We first proposed a quick linear inversion method based on the reduced instrument set for IST (LIST). Consequently, both the disharmony of linear relationship of instruments and the non-Markovian quantum correlations are detected and described with gauge freedom. However, because of the absence of constraints in the gauge optimization, the result of linear independent instruments is always the prior knowledge when the probability matrix is full rank. Moreover, the result of LIST is not guaranteed to be physical implementable. Then, a statistical method based on the maximum likelihood estimation for IST is proposed as MLE-IST with the capability utilizing overcomplete data. Based on the full instrument set, the MLE-IST tries to explicitly describe the detail of SE correlations {with adjustable and reduced dimensions. Based on the fact that dimensions of SE correlations in the current quantum devices are relatively low, dimensions of SE unitaries are reduced to the polynomial of time steps.} The results of MLE-IST is guaranteed to be physical implementable with constraints based on the assumptions of the quantum device. Specifically, we demonstrate how to implement IST on the current noisy quantum intermediate-scale quantum (NISQ) devices. The results of simulations and experiments exhibits significantly low SEP, which indicate the effectiveness of describing instruments and the non-Markovian quantum system including the initial state and the SE correlations. The IST provide an essential method for benchmarking and developing a quantum device under non-Markovian situation in the aspect of instrument set.

\section*{Acknowledgements}
This work is supported by Jiangsu Key R\&D Program Project BE2023011-2, Jiangsu Funding Program for Excellent Postdoctoral Talent No.2022ZB139, National Natural Science Foundation of China No.61960206005, and National Natural Science Foundation of China No.61871111.

\bibliographystyle{quantum}
\bibliography{citations}

\onecolumn\newpage
\appendix

\section{Example}\label{appendix:example}
\begin{example}[Non-uniqueness of SE unitaries and initial SE state]\label{eg:nunique} The non-uniqueness can be verified by the fact that
\begin{align}
  &\mathcal{U}_{t:t+1} \circ \mathcal{A}^{(t)}_{x_t} \circ \mathcal{U}_{t-1:1} \circ \mathcal{A}^{(t-1)}_{x_{t-1}} \left(\rho_{SE}^{(t-1)}\right) \\
  =& \mathcal{U}_{t:t+1} \circ (I\otimes \mathcal{U}_E) \circ\mathcal{A}^{(t)}_{x_t} \circ (I\otimes \mathcal{U}_E^{\dagger}) \circ\mathcal{U}_{t-1:1} \circ{(I\otimes \mathcal{V}_E)} \circ \mathcal{A}^{(t-1)}_{x_{t-1}} \circ{(I\otimes \mathcal{V}_E^{\dagger})}\left(\rho_{SE}^{(t-1)}\right)\\
  =&\mathcal{U}^{\prime}_{t:t+1} \circ \mathcal{A}^{(t)}_{x_t} \circ \mathcal{U}^{\prime}_{t-1:1} \circ \mathcal{A}^{(t-1)}_{x_{t-1}} \left(\rho_{SE}^{(t-1){\prime}}\right),
\end{align}
where $\mathcal{U}_E$ and $\mathcal{V}_E$ are unitaries act on the environment,
\begin{align}
  \mathcal{U}^{\prime}_{t:t+1} &= \mathcal{U}_{t:t+1} \circ (I\otimes \mathcal{U}_E),\\
  \mathcal{U}^{\prime}_{t-1:1} &= (I\otimes \mathcal{U}_E^{\dagger}) \circ\mathcal{U}_{t-1:1} \circ{(I\otimes \mathcal{V}_E)}, \\
  \rho_{SE}^{(t-1){\prime}} &= {(I\otimes \mathcal{V}_E^{\dagger})}\left(\rho_{SE}^{(t-1)}\right),
\end{align}
which means both $\left\{\mathcal{U}_{t:t+1}, \mathcal{U}_{t-1:t}, \rho_{SE}^{(t-1)}\right\}$ and $\left\{\mathcal{U}^{\prime}_{t:t+1}, \mathcal{U}^{\prime}_{t-1:t}, \rho_{SE}^{(t-1)\prime}\right\}$ are feasible to the probability.
\end{example}

\begin{example}[Non-uniqueness of process tensor and instruments]\label{eg:nunique_reduced} 
  The non-uniqueness of process tensor and instruments can be verified by the fact that
  \begin{align}
    &\Upsilon_{\mathcal{T}}^\dagger\left(A^{(0)}_{x_0}\otimes\dots\otimes A^{(k-1)}_{x_{k-1}}\right) \\
    =& \Upsilon_{\mathcal{T}}^\dagger\left(B^{(0)}\otimes\dots\otimes B^{(k-1)}\right)^{-1}\left(B^{(0)}\otimes\dots\otimes B^{(k-1)}\right)\left(A^{(0)}_{x_0}\otimes\dots\otimes A^{(k-1)}_{x_{k-1}}\right)\\
    =&(\Upsilon_{\mathcal{T}}^\prime)^\dagger\left(A^{(0)\prime}_{x_0}\otimes\dots\otimes A^{(k-1)\prime}_{x_{k-1}}\right),
  \end{align}
  where $A^{(t)}_{x_{t}}$ and $B^{(t)}A^{(t)}_{x_{t}}$ are PTMs of CPTNI instruments,
  \begin{align}
    (\Upsilon_{\mathcal{T}}^\prime)^\dagger &= \Upsilon_{\mathcal{T}}^\dagger\left(B^{(0)}\otimes\dots\otimes B^{(k-1)}\right)^{-1}\\
    A^{(t)\prime}_{x_{t}} &= B^{(t)}A^{(t)}_{x_{t}}, \forall t
  \end{align}
    which means both $\left\{\Upsilon_{\mathcal{T}}, \mathcal{J} := \left\{\left\{A^{(t)}_{x_{t}}\right\}\right\}_{t=0}^{k-1}\right\}$ and $\left\{\Upsilon^\prime_{\mathcal{T}},\mathcal{J}^{\prime} := \left\{\left\{A^{(t)\prime}_{x_{t}}\right\}\right\}_{t=0}^{k-1}\right\}$ are feasible to the probability.
  \end{example}

\section{Likelihood Function}\label{appendix:likelihood}

Specifying a sequence $\bm{x}$, the probability defined in Eq.~\eqref{eq:se_evo_prob_tr} is measured by repeating the experiment $n_s$ times and recording $n_{\bm{x}}$, which denotes how many times the desired outputs occur. Therefore, we use the general likelihood function of instrument set
\begin{align}
  \mathcal{L}(\hat{\mathfrak{I}}) = \prod_{\bm{x}}(\hat{p}_{\bm{x}})^{n_{\bm{x}}}(1-\hat{p}_{\bm{x}})^{n_{s}-n_{\bm{x}}},
\end{align}
where $\hat{p}_{\bm{x}}$ is the probability estimator modeled by parameters.

By exploiting the central limit theorem, each term of the likelihood can be rewritten as a normal distribution,
\begin{align}
  \mathcal{L}(\hat{\mathfrak{I}}) = \prod_{\bm{x}}\exp\left[-\frac{\left(\tilde{p}_{\bm{x}}-\hat{p}_{\bm{x}}\right)^2}{\sigma_{\bm{x}}^2}\right],
\end{align}
where $\tilde{p}_{\bm{x}}=n_{bm{x}}/n_s$ represents the measured probability, $\sigma_{\bm{x}}^2=\tilde{p}_{\bm{x}}(1-\tilde{p}_{\bm{x}})/n_s$ is the sampling variance in the measurement $m_{\bm{x}}$. Exploiting the the monotonic logarithm function, maximizing $\mathcal{L}$ is equivalent to minimizing the weighted mean square error (MSE)
\begin{align}
  l(\hat{\mathfrak{I}})=& -\log(\mathcal{L}(\hat{\mathfrak{I}})) = \sum_{\bm{x}}\frac{\left(\tilde{p}_{\bm{x}}-\hat{p}_{\bm{x}}\right)^2}{\sigma_{\bm{x}}^2}.
\end{align}

\section{MLE-IST with reduced instrument set}\label{appendix:reduced_mle}

While performing MLE-IST with reduced instrument set (reduced MLE-IST), an instrument $\mathcal{A}_{x_t}^{(t)}$ is modeled as it in the full instrument set MLE-IST with CPTNI constraints. The process tensor is modeled by $\hat{\Upsilon}_\mathcal{T} \in [-1,1]^{d^{2k}\times d^{2k}}$ with CP and causality constraints. Utilizing the CJI representation of process tensor $\hat{\Gamma}_\mathcal{T}$, CP requires positive semidefiniteness of $\hat{\Gamma}_\mathcal{T}$, while the causality requires $\langle\!\langle \hat{\Gamma} \vert 0\rangle\!\rangle = 1$ and
\begin{gather}
  \langle\!\langle \hat{\Gamma} \vert P_{\mathrm{ban}}\rangle\!\rangle = 0, ~\forall P_{\mathrm{ban}} := I^{\otimes{2t+1}} \otimes \left(\tilde{Q}_{2t+2} \otimes Q_{2t+3}\otimes\dots\otimes Q_{2k-1}\right), \forall t,\label{eq:proc_tensor_causality}\\
  \tilde{Q} \in \left\{P_1,\dots,P_{d^2-1}\right\},\\
  Q \in \left\{P_0,P_1,\dots,P_{d^2-1}\right\}.
\end{gather}
Then, the estimator is modeled by
\begin{align}
  \hat{p}_{\bm{x}}=\mathrm{Tr}\left[\hat{\Upsilon}_{\mathcal{T}}^\dagger\left(\hat{R}^{(0)}_{x_0}\otimes\dots\otimes \hat{R}^{(k-1)}_{x_{k-1}}\right)\right].
\end{align}

Consequently, the optimization problem of the reduced MLE-IST is given by
\begin{align}\label{eq:mle_ist_opt_prob_reduced}
  \min&_{\substack{\hat{\Upsilon}_\mathcal{T}, \hat{R}^{(t)}_{x_t},\forall x_t, t
  }}~l(\hat{\mathfrak{I}})\\
  s.t.~& \eqref{eq:full_def_cp_costraint}, \eqref{eq:full_def_tni_costraint}, \eqref{eq:full_def_ptm_val_constraint}, \eqref{eq:proc_tensor_causality}, \notag\\
  & \langle\!\langle \hat{\Gamma} \vert 0\rangle\!\rangle = 1. \tag{C10} \label{eq:reduced_def_causality_2}
\end{align}
The reduced MLE-IST requires $\sum_{t=0}^{k-1}m_t d^4 + d^{4k} - \frac{d^{2k}-d^2}{d+1}$ parameters to estimate the reduced instrument set.

Obviously, this formulation constrains the instrument set to a relatively simple form that all inaccessible initial state and SE unitary dynamics are modeled in a vector represented process tensor. However, it requires $\mathcal{O}(d^{4k})$ parameters which increase exponentially with respect to the non-Markovian order $k$. It is intractable to solve the problem with exponentially increasing number of parameters. Therefore, we propose the reduced instrument set MLE-IST framework but have not implemented it for simulations and experiments.

\end{document}